\documentclass[12pt]{iopart}
\usepackage[english]{babel}
\usepackage[utf8]{inputenc}
\usepackage{graphicx,epstopdf}
\usepackage[colorinlistoftodos]{todonotes}
\usepackage{epsfig,color,textcase}
\usepackage{hyperref}
\usepackage{doi}
\begin{document}

\title{Correlations induced orbital ordering and cooperative Jahn-Teller distortion in the paramagnetic insulator KCrF$_3$}
\author{D. Novoselov$^{1,2}$, Dm.M. Korotin$^{1}$, V.I. Anisimov$^{1,2}$}

\address{$^1$ Institute of Metal Physics, S.Kovalevskoy St. 18, 620990 Yekaterinburg, Russia}
\address{$^2$ Ural Federal University, Mira St. 19, 620002 Yekaterinburg, Russia}
\ead{novoselov@imp.uran.ru}
\begin{abstract}
We investigate the origin of the orbital ordering in the paramagnetic phase of KCrF$_3$. All previous studies described structural parameters of the paramagnetic phase using a magnetic ordering in the compound. 
Our simulations of real paramagnetic KCrF$_3$ were performed within an approach combining density functional theory and dynamical mean field theory (DFT+DMFT).
As a result, it was found that the experimentally observed cooperative Jahn-Teller effect is successfully described in a lattice relaxation calculation for structure without any long-range magnetic ordering. It is established that the existence of the orbital ordering even in undistorted perovskite structure clearly confirms the electronic origin of the orbital ordering in KCrF$_3$.
\end{abstract}


\section{Introduction}

KCrF$_3$ is a compound with electronic, magnetic and structural degrees of freedom coupled with each other, and all of them play important roles for its physical properties. 
This compound has four structural phases (in different temperature)~\cite{Margadonna2006}: cubic perovskite, tetragonal structure with slightly compressed/expanded CrF$_6$ octahedra, monoclinic structure with distorted and tilted CrF$_6$ octahedra and monoclinic phase without the octahedra tilting.

The transition metal ion in KCrF$_3$ has a quarter-filled degenerate $d$-states of the $e_g$ symmetry that assumes the existence of cooperative Jahn-Teller distortions (CJTD) of surrounding fluorine octahedra. CJTD is accompanied by orbital ordering of the occupied $d$-orbitals of the neighboring Cr ions. One can find an analogy between KCrF$_3$ and well known correlated materials LaMnO$_3$~\cite{Medvedeva2002} and KCuF$_3$~\cite{Leonov2008} with similar crystal structure and orbital ordering.

Since electronic, orbital and magnetic degrees of freedom are interconnected, the source of the orbital ordering in KCrF$_3$ is unclear. Does the distortion of CrF$_6$ octahedra enforce an electron to occupy a specific orbital or Coulomb correlations are strong enough to destroy the degeneracy of the $d-e_g$ energy bands and to fill distinct electronic state? Several works were devoted to this questions, see e.g. \cite{2012SurSc.606.1422W} and \cite{Xu2008}. Authors of these papers used the same approach -- DFT+U for the electronic structure description and came to different conclusions about the origin of orbital ordering.

In KCrF$_3$, the chromium ion $d$-shell configuration is $t^3_{2g}e^1_g$. The existence of the partially filled energy band means the necessity to take into account Coulomb correlations for a proper electronic and crystal structures description for KCrF$_3$. It can be done, for example, within DFT+U~\cite{Anisimov1991,Anisimov1991a} or DFT+DMFT~\cite{PhysRevLett.62.324,RevModPhys.68.13, RevModPhys.78.865, Anisimov1997, PhysRevB.57.6884} approaches. The former approach imposes the existence of a long-range magnetic ordering in crystal. Apparently, KCrF$_3$ has N\'eel temperature $T_N = 26~K$ and two structural transitions from the cubic to tetragonal and from tetragonal to monoclinic phases observed at 923~$K$ and 250~$K$, respectively~\cite{PhysRevB.77.075113}.
But the employment of the DFT+U approach for describing non-magnetic phases is not justified due to the fact that it assumes the presence of a magnetic order.

Recently, structural and electronic properties of the similar compound KCuF$_3$ were successfully described~\cite{Leonov2008} within the DFT+DMFT approach. This method takes into account dynamic electronic correlations and doesn't put a constrain onto magnetic structure of a system. Therefore, the paramagnetic tetragonal phase of KCrF$_3$ could be treated employing a more appropriate method. 

We use the DFT+DMFT approach to simulate CJTD in KCrF$_3$ to obtain the ground state crystal structure parameters and to answer the question -- is it possible to found the orbital ordering in KCrF$_3$ without the existence of the Jahn-Teller distortion in the crystal?

\section{Method}

\begin{figure}[!ht]
\begin{minipage}[h]{0.32\linewidth}
\center{\includegraphics[width=0.95\linewidth]{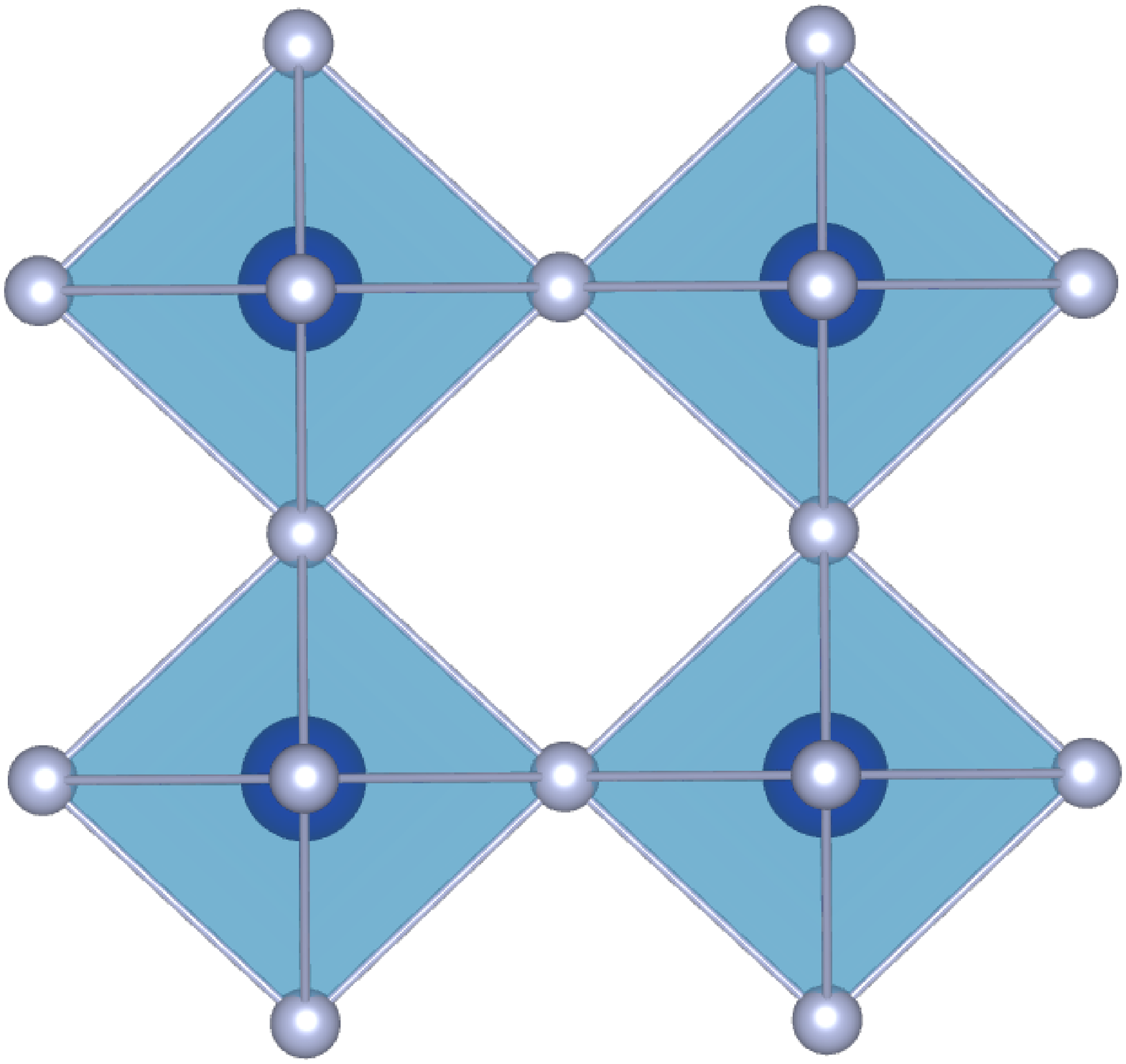} \\ a)}
\end{minipage}
\hfill
\begin{minipage}[h]{0.32\linewidth}
\center{\includegraphics[width=1.0\linewidth]{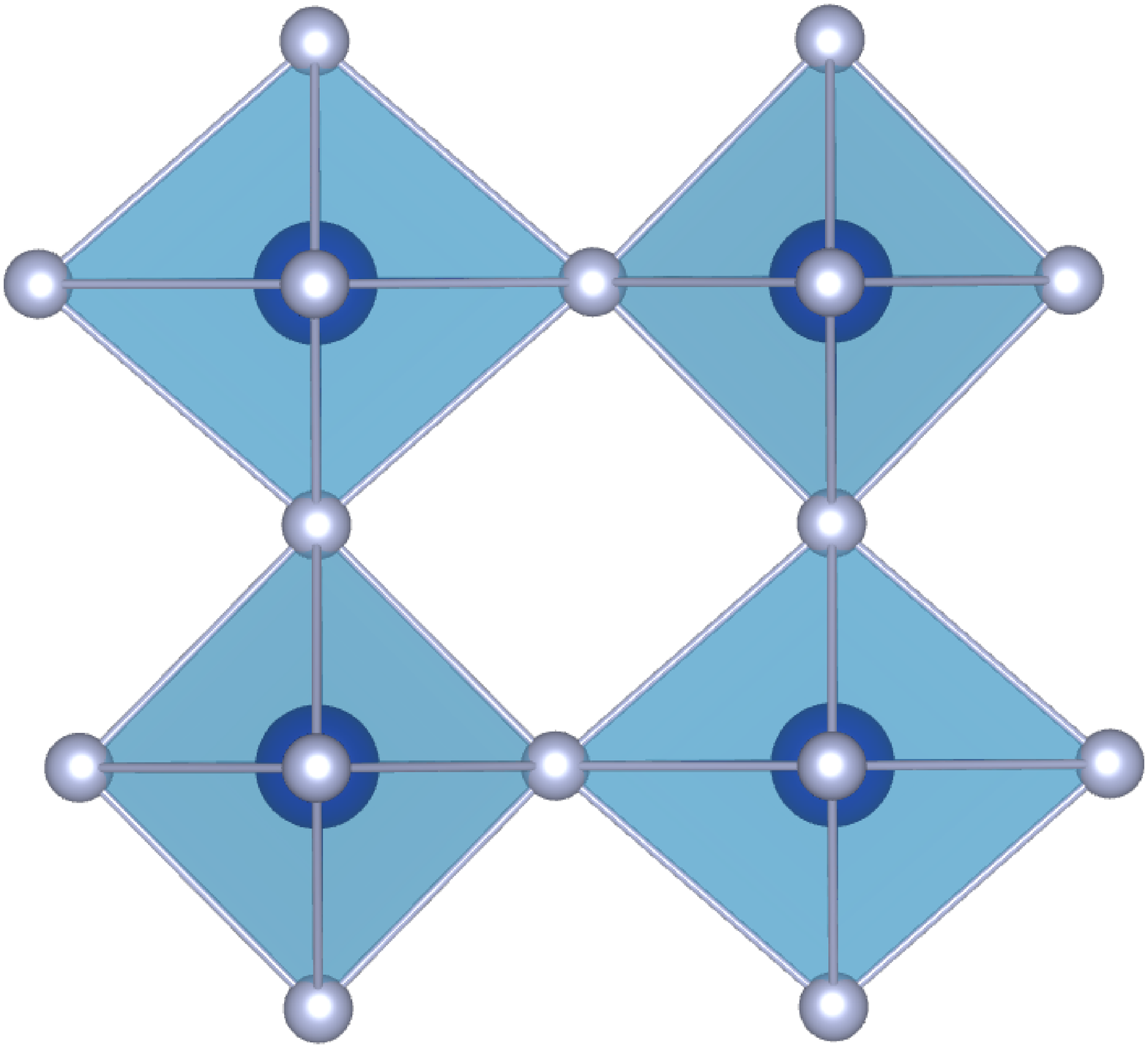} \\ b)}
\end{minipage}
\hfill
\begin{minipage}[h]{0.32\linewidth}
\center{\includegraphics[width=1.0\linewidth]{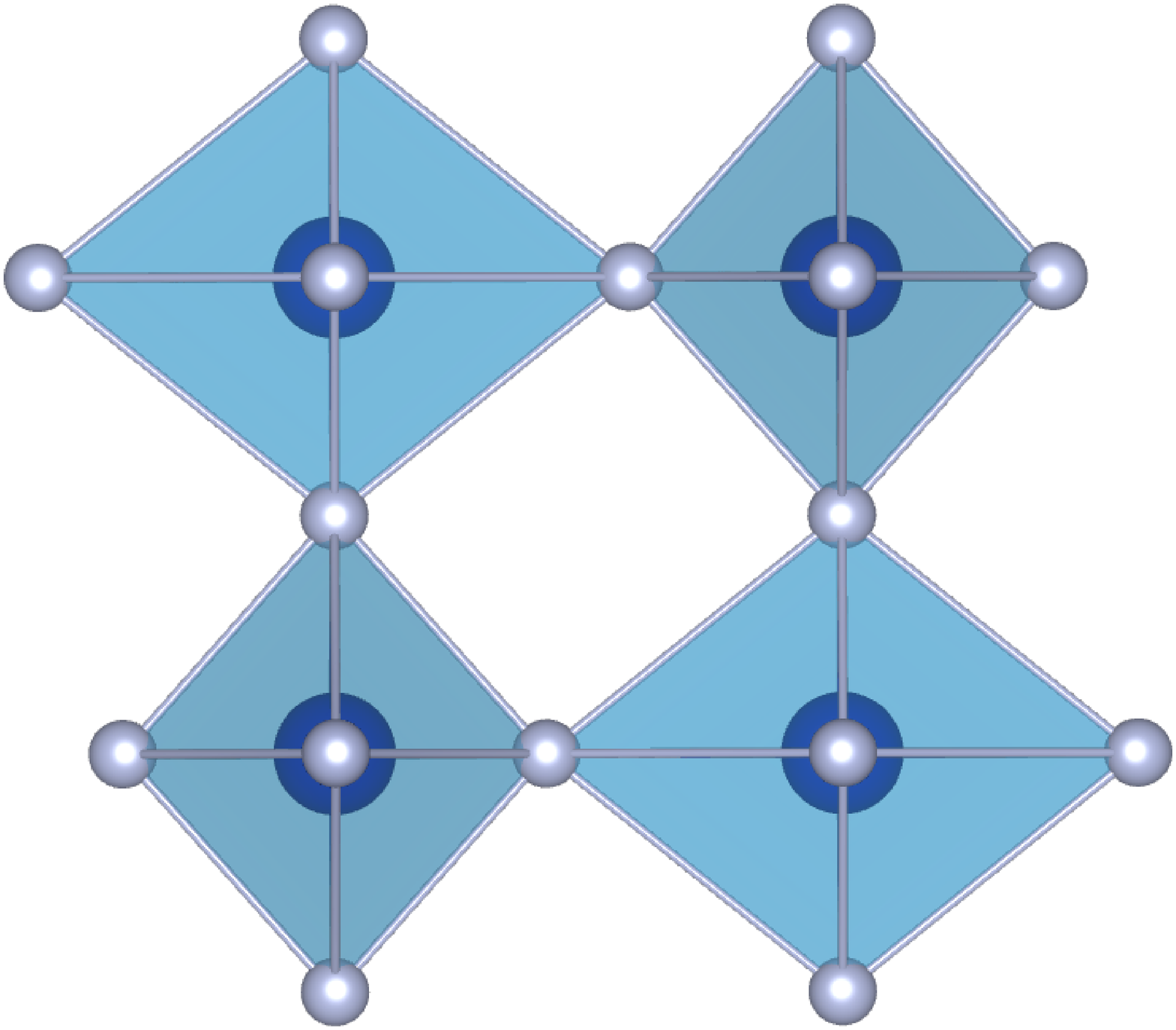} \\ c)}
\end{minipage}
\caption{(color online) Crystal structure of paramagnetic KCrF$_3$ for different Jahn-teller distortions in the $\bf{xy}$-plane: a) $\delta_{JT}=0.0\%$ b) $\delta_{JT}=2.7\%$ c) $\delta_{JT}=5.85\%$.
The fluorine ions (small light spheres) form octahedra around the Cr ions (dark large spheres). The K ions are not shown.} 
\label{fig:structure}
\end{figure}

For a thorough description of the interplay between the electronic and  structural degrees of freedom we perform a series self-consistent {\em ab-initio} calculations for various degrees of Jahn-Teller distortions $\delta_{JT}$ in the interval from 0 to 6\%. 
The evolution of the CrF$_6$ octahedra namely compression/expansion from the cubic perovskite structure to the maximally distorted tetragonal phase is shown schematically in figure \ref{fig:structure}. The CJTD leads to an elongation of the CrF$_6$ octahedra along the $x$ and $y$ axes and an antiferro-distortive pattern in the $xy$ plane.
Only a single internal structure parameter, namely, the shift of the in-plane fluorine atom from the Cr-Cr bond center is needed to describe the lattice distortion.

For all studied crystal structures two calculation stages are fulfilled. 
The first stage consists of a construction of a noncorrelated model Hamiltonian for the partially filled chromium ion $d$-states. A self-consistent DFT calculation within GGA approximation is performed for the chosen crystal structure with Quantum-ESPRESSO package~\cite{Giannozzi2009}. This package implements ultrasoft pseudopotential formalism with plane waves basis set. The cutoff energy for plane waves basis is set to 50 Ry. Integration in the reciprocal space is done on a regular 12x12x12 $k$-points mesh. Then a basis of atomic-centered Wannier functions for correlated states description is constructed~\cite{Korotin2008}. Wannier functions are generated as a projection of atomic Cr-$d$ wavefunctions onto the Bloch states with energies close to the Fermi level. The resulting basis functions consist of 97.7\% of the Cr-$d$ orbitals with a light admixture of the nearest F-$p$ states. The obtained Hamiltonian matrix in the Wannier functions basis describes the partially filled Cr-$d$ states within the DFT approach, and it is used as input for the DMFT calculation.

The second stage of distorted structure simulation is the DMFT calculation. The Hubbard model is solved with  continuous-time quantum Monte-Carlo (CT-QMC) solver~\cite{Werner2006a,Gull} for the Hamiltonian obtained on the previous stage. 
The Coulomb interaction matrix is constructed using the Hubbard $U = 6~eV$ and Hund $J = 0.88~eV$ parameter values~\cite{PhysRevB.77.075113}.
We use the DMFT approach realization implemented in the AMULET package~\cite{amulet}. 

The total energy of the cell is calculated as:
\begin{eqnarray}
E_{tot} = E_{DFT} - E^{Kin}_{DFT} + E^{Kin}_{DMFT} - N_{at} \cdot ( E_{DC} - U<nn> ),
\label{Etot}
\end{eqnarray}
where $E_{DFT}$ -- total energy from the DFT calculation, $E^{Kin}_{DFT}$ -- DFT kinetic energy, $E^{Kin}_{DMFT}$ -- DMFT kinetic energy, $N_{at}$ -- number of atoms, $E_{DC}$ -- double counting energy, $U<nn>$ -- Coulomb energy.
To determine an equilibrium crystal structure the forces distorting the CrF$_6$ octahedra are calculated and crystal structure is relaxed down to zero total force.
The total force acting on the fluorine ions is calculated as a derivative of the total energy over ion displacement between two Cr ions.

\section{Results and discussion}
\begin{figure}[!ht]
\centerline{\includegraphics[width=0.5\columnwidth,clip]{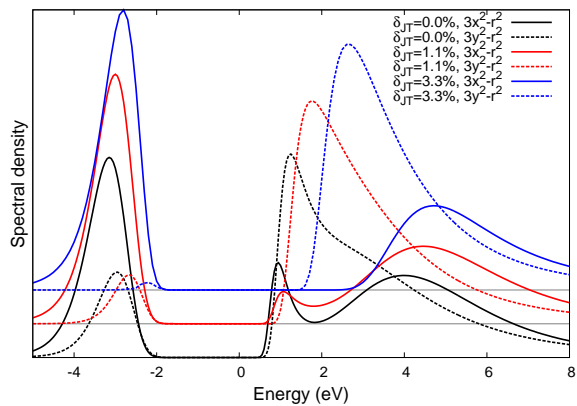}}
\caption{(color online) Orbital resolved spectral densities of the Cr $e_g$ states in the paramagnetic KCrF$_3$ obtained within the DFT+DMFT method for different JT distortions. Fermi level is placed at zero energy.} 
\label{fig:dos}
\end{figure}

As a reference crystal structure we used parameters from~\cite{Xiao2010} with variations in octahedra distortion degree.
The spectral densities of the Cr $e_g$ orbitals for various JT distortions calculated within the DFT+DMFT approach are shown in figure~\ref{fig:dos}. The insulating ground state was found for all distortion values under consideration. The bandwidth of the lower Hubbard band is about 2.5~eV and it doesn't depend on the crystal structure distortion. The energy band gap value varies from 2.4 to 3.4~eV.
Thus, the Cr $e_g$ spectra characteristics are in agreement with the previous results~\cite{PhysRevB.77.075113,PhysRevB.89.155109}.
It is noteworthy that the energy gap increases with the increase of the magnitude of the JT distortion. As a result of the distortion, the crystal field splits the $e_g$ energy levels. The Coulomb interaction correction enforces occupation of the lower $e_g$ band and increases the energy gap value.

\begin{figure}[!ht]
\centerline{\includegraphics[width=0.5\columnwidth,clip]{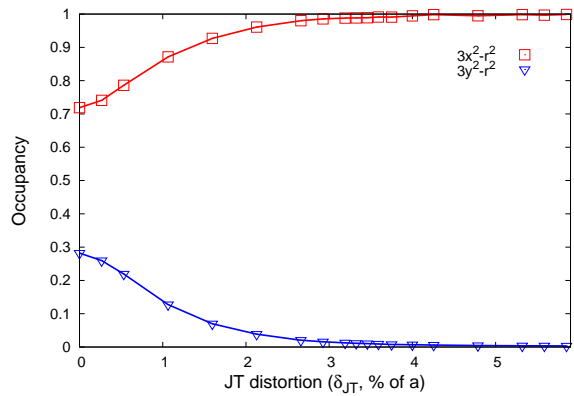}}
\caption{(color online) Dependence of the Cr $e_g$ orbitals occupancy in paramagnetic KCrF$_3$ computed within DFT+DMFT.} 
\label{fig:eg_occupancy}
\end{figure}

The lower Hubbard band for the distorted structure has the symmetry of the $3x^2-r^2$ orbital while the upper Hubbard band is mostly formed by  the electronic states of the $3y^2-r^2$ symmetry. With the reduction of the JT distortion, the $3x^2-r^2$ orbital contribution to the upper Hubbard band is increased.
It is clear from figure \ref{fig:dos} that for $\delta_{JT} = 3.3\%$ the $e_g$ states are fully polarized -- the spectral function of the $3x^2-r^2$ orbital is located below the Fermi level and the spectral function of $3y^2-r^2$ lies above. 
To approve this statement, the dependence of $e_g$ orbitals occupancy against the JT distortion is shown in figure~\ref{fig:eg_occupancy}.

It was shown in \cite{Wang2011} that for the undistorted cubic phase of KCrF$_3$ the GGA calculation results in almost degenerate $e_g$ states of the Cr ion. 
Therefore, it is impossible to obtain the orbital polarization picture in the cubic phase, at least without a direct consideration of strong electron-electron interactions. 
The similar situation seems to be true for KCuF$_3$~\cite{Leonov2008}. The orbital order parameter, defined as a difference between occupations of the two $e_g$ orbitals, is only about 12~\%, if one performs a calculation for the cubic perovskite structure of KCuF$_3$ without accounting of Coulomb correlations.

The direct inclusion of Coulomb correlations between the $d$-electrons of the Cr ions within DMFT leads to a different result. 
As one can see from figure~\ref{fig:eg_occupancy}, the orbital polarization exists even for the structure without the plane JT distortion. 
The orbital order parameter for the cubic phase is about 40\%.
Thus we can conclude that the presence of the electron-electron correlations leads to the orbital ordering in KCrF$_3$ even in the absence of the Jahn-Teller distortion.

Such a result is in disagreement with the overall finding of work \cite{Xu2014} where the authors conclude about the orbital ordering in KCrF$_3$ originating mostly from the CJTD effect. On the other hand, our results are in agreement with work \cite{Wang2011} where it was stated that electronic correlations play a significant role in the orbital polarization formation in KCrF$_3$.

\begin{figure}[!ht]
\centerline{\includegraphics[width=0.5\columnwidth,clip]{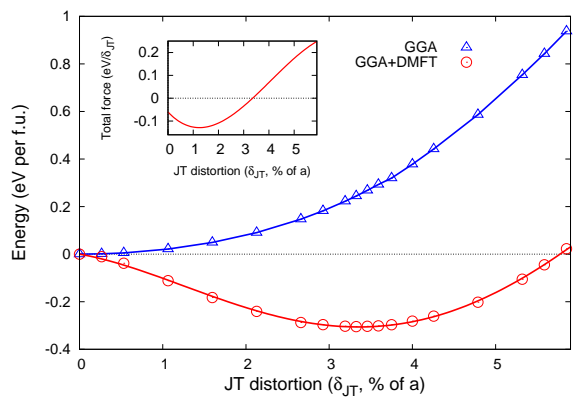}}
\caption{(color online) Total energy dependencies of the paramagnetic KCrF$_3$ over the Jahn-Teller distortion obtained within the DFT and DFT+DMFT methods.} 
\label{fig:total_energy}
\end{figure}

The result of the total energy calculations for several values of the plane JT distortion parameter $\delta_{JT}$ is shown in figure~\ref{fig:total_energy}. 
The total energy obtained within DFT is shown as a blue curve and the total energy obtained within the DFT+DMFT approach is shown as a red curve.
The total force calculated as a derivative of the total energy over the distortion is shown in the inset in figure \ref{fig:total_energy}.

It is clear that the DFT calculations do not allow one to reproduce the ground state tetragonal phase since the total energy dependence has a minimum for the cubic phase.
The total energy dependence obtained with strong electron-electron interaction taken into account has the minimum clearly for the distorted structure. The tetragonal phase is more than 1000~K lower than that of the cubic one in this energy range. The point where the total force becomes zero and the resulting total energy passes through the minimum corresponds to the JT distortion equal to about 3.3~\% which is close to the experimentally observed 2.7~\%.

This result indicates that the Jahn-Teller distortion in the paramagnetic insulator KCrF$_3$ is predominantly caused by the electronic correlations.
Thus we can conclude that in this compound the Coulomb correlations result in the appearance of orbital order as well as the Jahn-Teller effect, which in its turn increases the splitting between the two $e_g$ levels leading eventually to the full orbital polarization.

\section{Conclusion}
We calculated the ground states crystal and electronic structures of paramagnetic phase of KCrF$_3$. The strong electronic correlations were described within DFT+DMFT which doesn't imply the existence of a long-range magnetic order. The significant orbital polarization of the $e_g$ states is obtained even for the undistorted perovskite structure. Also the necessity to take into account the Coulomb correlations to describe the cooperative Jahn-Teller distortion in the compound is shown. Therefore, we conclude that the strong electronic correlations lead to the occurrence of the orbital polarization effect that in its turn becomes a driving force for the Jahn-Teller distortion of the CrF$_6$ octahedra in KCrF$_3$.

\ack
The present work was supported by the grant of the Russian Science Foundation (project no. 14-22-00004).

\section*{References}
\bibliography{main}

\end{document}